\documentclass[twocolumn,letterpaper,aps,prd,superscriptaddress,longbibliography,floatfix]{revtex4-2}
\usepackage{graphicx}	% Include figure files
\usepackage{xspace}     % Include xspace

\begin{document}

%%%%%%%%%%%%%%%%%%%%%%%%%%%%%%%%%%%%%%%%%%%%%%%%% Title of paper

\title{Transverse-single-spin asymmetries of charged pions at 
midrapidity in transversely polarized $p$$+$$p$ collisions at 
$\sqrt{s}=200$ GeV }

\newcommand{\abilene}{Abilene Christian University, Abilene, Texas 79699, USA}
\newcommand{\augie}{Department of Physics, Augustana University, Sioux Falls, South Dakota 57197, USA}
\newcommand{\banaras}{Department of Physics, Banaras Hindu University, Varanasi 221005, India}
\newcommand{\barc}{Bhabha Atomic Research Centre, Bombay 400 085, India}
\newcommand{\baruch}{Baruch College, City University of New York, New York, New York, 10010 USA}
\newcommand{\bnlcoll}{Collider-Accelerator Department, Brookhaven National Laboratory, Upton, New York 11973-5000, USA}
\newcommand{\bnlphys}{Physics Department, Brookhaven National Laboratory, Upton, New York 11973-5000, USA}
\newcommand{\caucr}{University of California-Riverside, Riverside, California 92521, USA}
\newcommand{\charlesczech}{Charles University, Ovocn\'{y} trh 5, Praha 1, 116 36, Prague, Czech Republic}
\newcommand{\cns}{Center for Nuclear Study, Graduate School of Science, University of Tokyo, 7-3-1 Hongo, Bunkyo, Tokyo 113-0033, Japan}
\newcommand{\colorado}{University of Colorado, Boulder, Colorado 80309, USA}
\newcommand{\columbia}{Columbia University, New York, New York 10027 and Nevis Laboratories, Irvington, New York 10533, USA}
\newcommand{\czechtech}{Czech Technical University, Zikova 4, 166 36 Prague 6, Czech Republic}
\newcommand{\debrecen}{Debrecen University, H-4010 Debrecen, Egyetem t{\'e}r 1, Hungary}
\newcommand{\elte}{ELTE, E{\"o}tv{\"o}s Lor{\'a}nd University, H-1117 Budapest, P{\'a}zm{\'a}ny P.~s.~1/A, Hungary}
\newcommand{\eszterhazy}{Eszterh\'azy K\'aroly University, K\'aroly R\'obert Campus, H-3200 Gy\"ongy\"os, M\'atrai \'ut 36, Hungary}
\newcommand{\ewha}{Ewha Womans University, Seoul 120-750, Korea}
\newcommand{\famu}{Florida A\&M University, Tallahassee, FL 32307, USA}
\newcommand{\fsu}{Florida State University, Tallahassee, Florida 32306, USA}
\newcommand{\gsu}{Georgia State University, Atlanta, Georgia 30303, USA}
\newcommand{\hiroshima}{Hiroshima University, Kagamiyama, Higashi-Hiroshima 739-8526, Japan}
\newcommand{\howard}{Department of Physics and Astronomy, Howard University, Washington, DC 20059, USA}
\newcommand{\ihepprot}{IHEP Protvino, State Research Center of Russian Federation, Institute for High Energy Physics, Protvino, 142281, Russia}
\newcommand{\illuiuc}{University of Illinois at Urbana-Champaign, Urbana, Illinois 61801, USA}
\newcommand{\inrras}{Institute for Nuclear Research of the Russian Academy of Sciences, prospekt 60-letiya Oktyabrya 7a, Moscow 117312, Russia}
\newcommand{\instpasczech}{Institute of Physics, Academy of Sciences of the Czech Republic, Na Slovance 2, 182 21 Prague 8, Czech Republic}
\newcommand{\isu}{Iowa State University, Ames, Iowa 50011, USA}
\newcommand{\jaea}{Advanced Science Research Center, Japan Atomic Energy Agency, 2-4 Shirakata Shirane, Tokai-mura, Naka-gun, Ibaraki-ken 319-1195, Japan}
\newcommand{\jeonbuk}{Jeonbuk National University, Jeonju, 54896, Korea}
\newcommand{\kek}{KEK, High Energy Accelerator Research Organization, Tsukuba, Ibaraki 305-0801, Japan}
\newcommand{\korea}{Korea University, Seoul 02841, Korea}
\newcommand{\kurchatov}{National Research Center ``Kurchatov Institute", Moscow, 123098 Russia}
\newcommand{\kyoto}{Kyoto University, Kyoto 606-8502, Japan}
\newcommand{\lawllnl}{Lawrence Livermore National Laboratory, Livermore, California 94550, USA}
\newcommand{\losalamos}{Los Alamos National Laboratory, Los Alamos, New Mexico 87545, USA}
\newcommand{\lund}{Department of Physics, Lund University, Box 118, SE-221 00 Lund, Sweden}
\newcommand{\lyon}{IPNL, CNRS/IN2P3, Univ Lyon, Université Lyon 1, F-69622, Villeurbanne, France}
\newcommand{\maryland}{University of Maryland, College Park, Maryland 20742, USA}
\newcommand{\mass}{Department of Physics, University of Massachusetts, Amherst, Massachusetts 01003-9337, USA}
\newcommand{\michigan}{Department of Physics, University of Michigan, Ann Arbor, Michigan 48109-1040, USA}
\newcommand{\muhlenberg}{Muhlenberg College, Allentown, Pennsylvania 18104-5586, USA}
\newcommand{\nara}{Nara Women's University, Kita-uoya Nishi-machi Nara 630-8506, Japan}
\newcommand{\natmephi}{National Research Nuclear University, MEPhI, Moscow Engineering Physics Institute, Moscow, 115409, Russia}
\newcommand{\newmex}{University of New Mexico, Albuquerque, New Mexico 87131, USA}
\newcommand{\nmsu}{New Mexico State University, Las Cruces, New Mexico 88003, USA}
\newcommand{\northcg}{Physics and Astronomy Department, University of North Carolina at Greensboro, Greensboro, North Carolina 27412, USA}
\newcommand{\ohio}{Department of Physics and Astronomy, Ohio University, Athens, Ohio 45701, USA}
\newcommand{\ornl}{Oak Ridge National Laboratory, Oak Ridge, Tennessee 37831, USA}
\newcommand{\orsay}{IPN-Orsay, Univ.~Paris-Sud, CNRS/IN2P3, Universit\'e Paris-Saclay, BP1, F-91406, Orsay, France}
\newcommand{\peking}{Peking University, Beijing 100871, People's Republic of China}
\newcommand{\pnpi}{PNPI, Petersburg Nuclear Physics Institute, Gatchina, Leningrad region, 188300, Russia}
\newcommand{\pusan}{Pusan National University, Pusan 46241, Korea}
\newcommand{\riken}{RIKEN Nishina Center for Accelerator-Based Science, Wako, Saitama 351-0198, Japan}
\newcommand{\rikjrbrc}{RIKEN BNL Research Center, Brookhaven National Laboratory, Upton, New York 11973-5000, USA}
\newcommand{\rikkyo}{Physics Department, Rikkyo University, 3-34-1 Nishi-Ikebukuro, Toshima, Tokyo 171-8501, Japan}
\newcommand{\saispbstu}{Saint Petersburg State Polytechnic University, St.~Petersburg, 195251 Russia}
\newcommand{\seoulnat}{Department of Physics and Astronomy, Seoul National University, Seoul 151-742, Korea}
\newcommand{\stonybrkc}{Chemistry Department, Stony Brook University, SUNY, Stony Brook, New York 11794-3400, USA}
\newcommand{\stonycrkp}{Department of Physics and Astronomy, Stony Brook University, SUNY, Stony Brook, New York 11794-3800, USA}
\newcommand{\tenn}{University of Tennessee, Knoxville, Tennessee 37996, USA}
\newcommand{\texsu}{Texas Southern University, Houston, TX 77004, USA}
\newcommand{\titech}{Department of Physics, Tokyo Institute of Technology, Oh-okayama, Meguro, Tokyo 152-8551, Japan}
\newcommand{\tsukuba}{Tomonaga Center for the History of the Universe, University of Tsukuba, Tsukuba, Ibaraki 305, Japan}
\newcommand{\vandy}{Vanderbilt University, Nashville, Tennessee 37235, USA}
\newcommand{\weizmann}{Weizmann Institute, Rehovot 76100, Israel}
\newcommand{\wigner}{Institute for Particle and Nuclear Physics, Wigner Research Centre for Physics, Hungarian Academy of Sciences (Wigner RCP, RMKI) H-1525 Budapest 114, POBox 49, Budapest, Hungary}
\newcommand{\yonsei}{Yonsei University, IPAP, Seoul 120-749, Korea}
\newcommand{\zagreb}{Department of Physics, Faculty of Science, University of Zagreb, Bijeni\v{c}ka c.~32 HR-10002 Zagreb, Croatia}
\affiliation{\abilene}
\affiliation{\augie}
\affiliation{\banaras}
\affiliation{\barc}
\affiliation{\baruch}
\affiliation{\bnlcoll}
\affiliation{\bnlphys}
\affiliation{\caucr}
\affiliation{\charlesczech}
\affiliation{\cns}
\affiliation{\colorado}
\affiliation{\columbia}
\affiliation{\czechtech}
\affiliation{\debrecen}
\affiliation{\elte}
\affiliation{\eszterhazy}
\affiliation{\ewha}
\affiliation{\famu}
\affiliation{\fsu}
\affiliation{\gsu}
\affiliation{\hiroshima}
\affiliation{\howard}
\affiliation{\ihepprot}
\affiliation{\illuiuc}
\affiliation{\inrras}
\affiliation{\instpasczech}
\affiliation{\isu}
\affiliation{\jaea}
\affiliation{\jeonbuk}
\affiliation{\kek}
\affiliation{\korea}
\affiliation{\kurchatov}
\affiliation{\kyoto}
\affiliation{\lawllnl}
\affiliation{\losalamos}
\affiliation{\lund}
\affiliation{\lyon}
\affiliation{\maryland}
\affiliation{\mass}
\affiliation{\michigan}
\affiliation{\muhlenberg}
\affiliation{\nara}
\affiliation{\natmephi}
\affiliation{\newmex}
\affiliation{\nmsu}
\affiliation{\northcg}
\affiliation{\ohio}
\affiliation{\ornl}
\affiliation{\orsay}
\affiliation{\peking}
\affiliation{\pnpi}
\affiliation{\pusan}
\affiliation{\riken}
\affiliation{\rikjrbrc}
\affiliation{\rikkyo}
\affiliation{\saispbstu}
\affiliation{\seoulnat}
\affiliation{\stonybrkc}
\affiliation{\stonycrkp}
\affiliation{\tenn}
\affiliation{\texsu}
\affiliation{\titech}
\affiliation{\tsukuba}
\affiliation{\vandy}
\affiliation{\weizmann}
\affiliation{\wigner}
\affiliation{\yonsei}
\affiliation{\zagreb}
\author{U.A.~Acharya} \affiliation{\gsu} 
\author{C.~Aidala} \affiliation{\michigan} 
\author{Y.~Akiba} \email[PHENIX Spokesperson: ]{akiba@rcf.rhic.bnl.gov} \affiliation{\riken} \affiliation{\rikjrbrc} 
\author{M.~Alfred} \affiliation{\howard} 
\author{V.~Andrieux} \affiliation{\michigan} 
\author{N.~Apadula} \affiliation{\isu} 
\author{H.~Asano} \affiliation{\kyoto} \affiliation{\riken} 
\author{B.~Azmoun} \affiliation{\bnlphys} 
\author{V.~Babintsev} \affiliation{\ihepprot} 
\author{N.S.~Bandara} \affiliation{\mass} 
% ...
\collaboration{PHENIX Collaboration}  \noaffiliation

\date{\today}

%------------------------------------------------------------------------------|

\begin{abstract}

In 2015, the PHENIX collaboration has measured single-spin asymmetries 
for charged pions in transversely polarized $p$$+$$p$ collisions at 
the center of mass energy of $\sqrt{s}=200$ GeV. The pions were detected 
at central rapidities of $|\eta|<0.35$. The single-spin asymmetries are 
consistent with zero for each charge individually, as well as consistent 
with the previously published neutral-pion asymmetries in the same 
rapidity range. However, they show a slight indication of 
charge-dependent differences which may suggest a flavor dependence in 
the underlying mechanisms that create these asymmetries.

\end{abstract}

\maketitle

%\section{Info}
\section{Introduction}

Single-spin asymmetries have been measured in hadronic collisions over a 
wide range of energies for a number of final state 
particles~\cite{Adams:1991rw,STAR:2012ljf,PHENIX:2013wle,PHENIX:2021irw} 
and at various rapidity regions. Initially, calculations that were based 
entirely on perturbative quantum chromodynamics (pQCD) expected such 
asymmetries to be suppressed as they generally required a helicity-flip 
which would be proportional to the parton mass over the hard scale of 
the scattering process~\cite{Kane:1978nd}, though recent calculations 
suggest that perturbative contributions may be 
possible~\cite{Benic:2019zvg}. In contrast to the pQCD calculations, the 
measured single-spin asymmetries, $A_N$ for pion production in the 
forward direction were found to be quite 
sizeable~\cite{Adams:1991rw,STAR:2012ljf,PHENIX:2013wle}.

Initially, two different mechanisms were suggested to describe these 
asymmetries that attribute the effect to the nonperturbative parts of 
either the parton distribution functions~\cite{Sivers:1989cc} or the 
fragmentation functions~\cite{Collins:1992kk}. To do so, the 
traditional concept of parton distribution and fragmentation functions 
had to be extended to allow for explicit-transverse-momentum degrees of 
freedom. In the transverse-momentum-dependent (TMD) framework, it is 
however necessary that at least two scales are observed, the large 
scattering scale and a smaller scale related to the intrinsic transverse 
momenta of parton distribution and fragmentation. Both can be observed 
in semi-inclusive deeply inelastic scattering (SIDIS), as was first 
successfully demonstrated in~\cite{HERMES:2004mhh} for both suggested 
effects. 

Single-hadron final-state measurements in hadronic collisions have only 
one scale, typically given by the transverse momentum of the detected 
hadron. As such, a collinear framework that only relies on a single hard 
scale was suggested to describe these 
asymmetries~\cite{Efremov:1994dg,Qiu:1998ia}. In this framework, nonzero 
asymmetries require nonperturbative higher-twist correlations either in the 
initial state or the final state~\cite{Kanazawa:2000kp,Metz:2012ct}.  Subsequently, it 
was suggested that some of these correlation functions can be related to 
transverse-momentum moments of the TMD functions~\cite{Bacchetta:2008xw}, 
which unifies both approaches.  Thus, global fits of asymmetry measurements 
from SIDIS and hadron collisions have become possible and provide additional 
precision on the underlying functions of interest such as the quark 
transversity~\cite{Ralston:1979ys} or the Sivers~\cite{Sivers:1989cc} 
function in the proton. The latest of these comes 
from~\cite{Cammarota:2020qcw}, in which are found sizeable contributions 
from both initial and final state effects.  However, the final-state effects 
dominate in hadronic collisions where the measured asymmetries are large.

Pion transverse-single-spin asymmetries pick up both initial- and 
final-state effects.  While at forward rapidities, the asymmetries are quite 
sizeable in the energy range of the Relativistic Heavy Ion Collider (RHIC) 
at Brookhaven National Laboratory~\cite{STAR:2012ljf,PHENIX:2013wle}, 
neutral-pion and $\eta$ mesons have been shown to have vanishing asymmetries 
at central rapidities where the valence effects are expected to be less 
pronounced~\cite{PHENIX:2020mft}. In addition, in neutral-particle 
asymmetries, potential cancellations between different parton flavors could 
also cause these asymmetries to vanish. Charged pion asymmetries provide 
different flavor sensitivity via the fragmentation functions and could test 
whether such cancellations happen, as the dominant hard process at the 
transverse momenta of this measurement is given by quark-gluon scattering. 
The PHENIX detector at RHIC has measured charged pion single-spin 
asymmetries at central rapidities in transversely polarized $p$$+$$p$ 
collisions at a center-of-mass energy of $\sqrt{s}=200$ GeV.

In the following sections, the relevant detector systems and the 
accumulated data sets will be described. The documentation of the 
charged pion selection criteria, reconstruction efficiencies and various 
background corrections follow before evaluating the various systematic 
uncertainties for the asymmetry results which are then discussed.

\section{Data-sets}

In 2015, the PHENIX experiment accumulated 60 pb$^{-1}$ of transversely 
polarized $p$$+$$p$ collision data. As both beams were polarized 
with average polarizations of 60\% (counter-clockwise, yellow beam) and 
58\% (clockwise, blue beam)~\cite{polarimetry}, two independent 
measurements were initially performed using the spin information of one 
beam and averaging over the spin information of the second beam. After 
confirming the consistency of the single-spin asymmetries for each 
individual beam, the results were combined.

The PHENIX detector comprises a central Helmholtz double coil magnet 
that is surrounded by two central detector arms that cover 90 degrees 
azimuthally each, and a pseudorapidity range of $|\eta|< 0.35$. A 
detailed detector description can be found at~\cite{Adcox:2003zm}, while 
only the detector sub-systems relevant for this analysis will be 
presented in this paper. The central arms comprise drift (DC) and pad 
chambers (PC) at radii of 2.02--2.46 m (DC), 2.50 m (PC1), 4.00 m 
(PC2), and 5.00 m (PC3) that measure charged-particle 
momenta via the bending in the magnetic field. Between them, the gas 
ring-imaging \v{C}erenkov (RICH) detector is situated that helps 
identifying electrons, pions, and kaons at momentum thresholds of 0.03, 
4.7, and 16 GeV/$c$, respectively. Downstream of the PCs, two types of 
electromagnetic calorimeters, EMCal, are located. Six of eight sectors 
are instrumented with a lead-scintillator (PbSc) shashlik calorimeter while the 
remaining two sectors are instrumented with a lead-glass (PbGl)  calorimeter. 
The EMCal measures the energy of electrons and photons, as well as in 
part the energy of hadrons that start showering within the EMCal 
detector. The PbSc (PbGl) EMCal corresponds to 0.85 (1.05) nuclear 
interaction lengths. It also serves as a trigger detector for these 
particles, where typically three different energy thresholds are set up 
with different suppression factors in the data-taking.

Additionally, the beam-beam counters (BBCs), which comprise 64 quartz 
crystals and photo-multipliers each, located at the forward and backward 
rapidities of 3.1$ < |\eta| < $ 3.9 register hard collisions and the 
collision vertex position and the start time for time-of-flight 
measurements. The BBCs also serve to log the relative luminosities that 
have been accumulated in different spin states. Zero-degree-calorimeters 
at rapidities of $|\eta| > 6$ are used to cross check these relative 
luminosities as well as the correct beam spin orientation using the 
nonzero neutron single-spin asymmetries~\cite{PHENIX:2012tos}.

\section{Event and particle selection criteria}

Pion-candidate events were selected if the lowest threshold of the EMCal 
trigger condition was satisfied, which required that at least 1.4 GeV 
of energy was deposited by the candidate particle in any of 4 $\times$ 4 
EMCal towers. The BBC reconstructed event vertex had to be within 30 cm 
of the nominal beam interaction point along the beam axis.  Pions were 
then selected if their momentum was between 5 and 15 GeV/$c$ and the track 
matched the energy deposit in the EMCal. The track had to fire the RICH, 
and the shower shape in the EMCal had not to resemble that of an 
electromagnetic particle as given by a likelihood ratio between 
electromagnetic and other particles. To reduce the contamination by 
electrons in the pion data sample, the ratio of deposited energy, $E$, 
in the EMCal and reconstructed momentum, $p$, had to be $0.2 < E/p < 
0.8$. At low $E/p$ values, electrons that are produced in decays close 
to the EMCal and appear to have higher reconstructed momenta get 
rejected. While other electrons deposit all of their energy in the EMCal 
and typically get reconstructed at $E/p$ values around unity, pions only 
deposit a fraction of their energy in the EMCal and can thus be 
selected.

To estimate the asymmetries for electron background an electron-enhanced 
sample is selected in addition to the pion enhanced sample. For this 
sample the $E/p$ and shower shape selections are inverted and even more 
RICH activity was required. These two types of samples will be denoted 
as pion-enhanced and electron-enhanced samples.

\section{Asymmetry extraction}

The charged-pion yields were extracted starting by selecting them with 
the aforementioned criteria.

The single-spin asymmetries are then calculated for each beam, particle 
charge and signal type as:
\begin{equation}
    A_N = \frac{1}{P} \frac{\sqrt{N_L^\uparrow N_R^\downarrow} - \sqrt{N_L^\downarrow N_R^\uparrow}}{ \sqrt{N_L^\uparrow N_R^\downarrow} + \sqrt{N_L^\downarrow N_R^\uparrow}}\quad ,
    \label{eq:squareroot}
\end{equation}
where $N_{L/R}^{\uparrow/\downarrow}$ are the count rates to the left 
and right with respect to the polarized beam direction and spin 
orientation, for beams polarized up or down, respectively. $P$ is the 
average beam polarization. Additionally, as the detector sits not 
exactly at 90 degrees to the spin orientation, the asymmetry needs to be 
normalized with the average cosine of the azimuthal angle 
$\langle|\cos\phi|\rangle$ of all particle candidates $n$, $ 
\langle|\cos\phi|\rangle = \sum_i |\cos\phi_i|/n$, where $\phi_i$ is 
defined for each candidate $i$ relative to the spin orientation of the polarized beam, along 
its beam axis.

To evaluate the signal and background fractions in each sample, the raw 
candidate yields were corrected for the general reconstruction and 
acceptance efficiencies that were evaluated in single-particle 
simulations in {\sc geant}3~\cite{Brun:1994aa}. The trigger efficiencies 
were corrected by calculating the fraction of minimum-bias events, for 
which pion candidates were also triggered by the EMCal based trigger. 
The RICH efficiency was calculated similarly by comparing the ratio of 
pion candidates that require the RICH selection criterion over all 
candidates. Because electrons also fulfill the RICH requirements for lower 
transverse momenta, the RICH efficiencies were corrected for this 
background using {\sc pythia}6~\cite{Sjostrand:2001yu} simulations.

Given that pion candidates are triggered using an electromagnetic 
calorimeter, the overall reconstruction efficiencies are much lower than 
for neutral pions or electrons, see for example~\cite{PHENIX:2014axc} 
for more details on pion cross section measurements using PHENIX.

%-------------------------------------------- Fig_1
\begin{figure}[htb]
\includegraphics[width=1.0\linewidth]{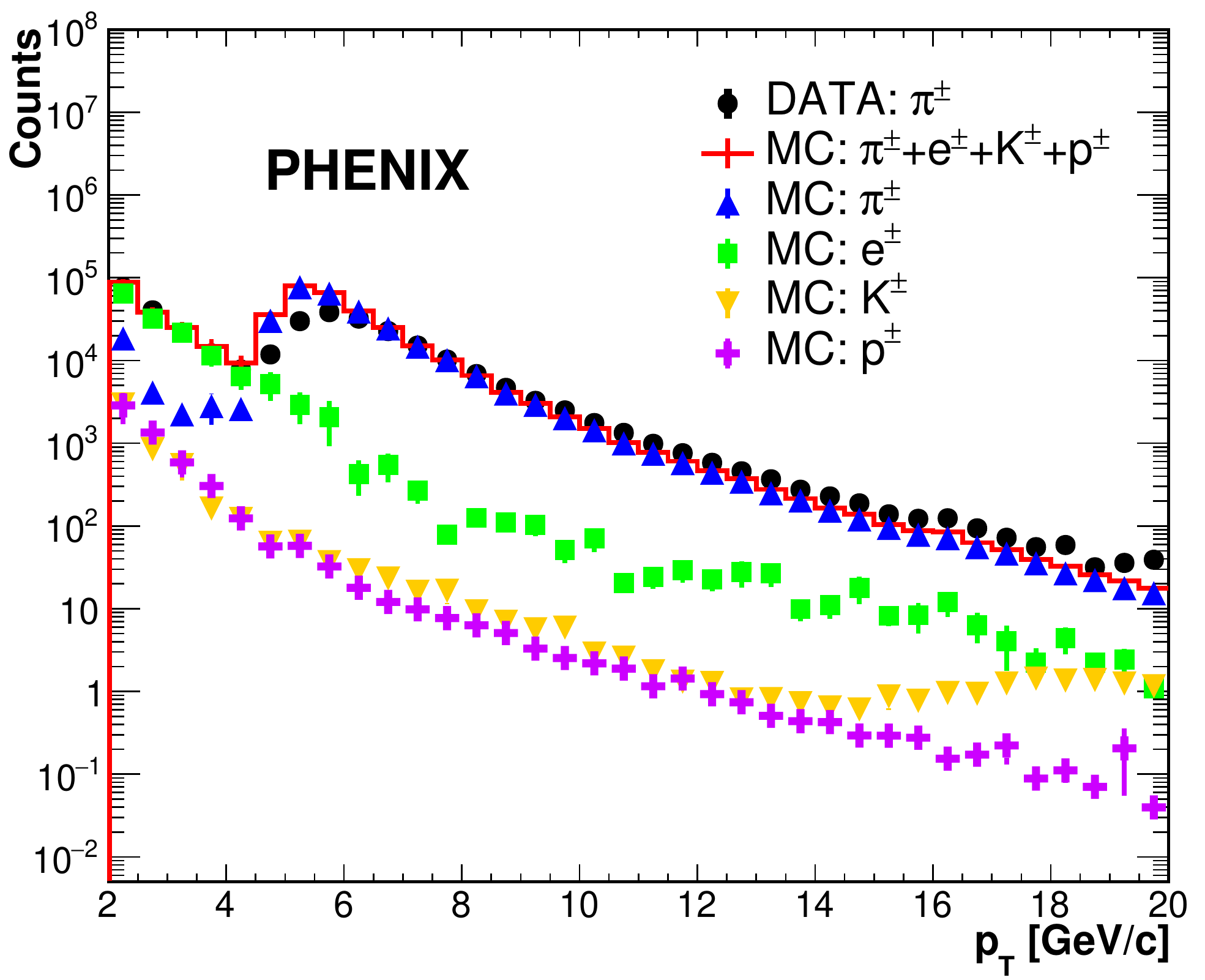}
\caption{Transverse-momentum distribution of reconstructed particles 
from data and MC simulation. }
\label{fig:pt_distribution}
\end{figure}

\begin{figure}[htb]
\includegraphics[width=1.0\linewidth]{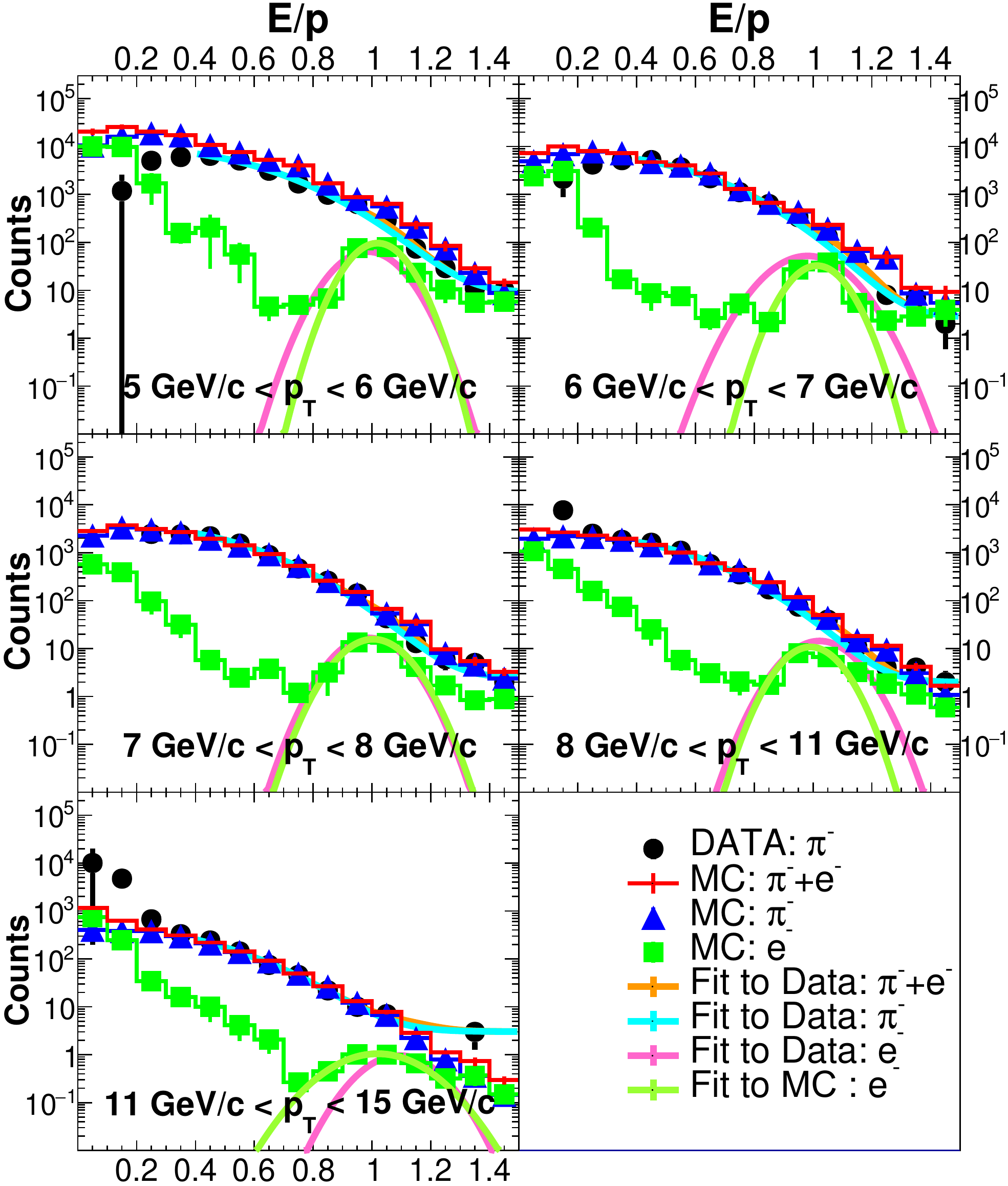}
\caption{Energy-over-momentum ratio distribution for negative pion and 
electron candidates from data and MC simulation for the pion-enhanced 
data sample.}
\label{fig:eoverp}
\end{figure}

The composition of the raw data yields as a function of the transverse 
momentum based on a {\sc pythia}6~\cite{Sjostrand:2001yu} Monte-Carlo (MC) 
simulation is shown in Fig.~\ref{fig:pt_distribution}, which shows that 
overall the data is well-described by the MC. The thresholds in the RICH for 
pions and kaons to emit \v{C}erenkov light are clearly visible at momenta of 
around 5 and 16 GeV/$c$, respectively.  The simulations, 
confirm that the selected sample is clearly dominated by pions, with 
electrons being the main background.  To calibrate the actual signal and 
background fractions in the pion and electron enhanced data samples, the 
$E/p$ distributions in both pion and electron enhanced data samples were fit 
by the shapes obtained from MC over an enlarged $E/p$ range. In particular, 
the electron contribution can be obtained by fitting a Gaussian close to 
unity which corresponds to the well-reconstructed electrons, that lose all 
their energy in the calorimeter, while the shape for the pions was directly 
taken from the simulation. The example for the raw yields and fits to the 
pion-enhanced data sample can be seen in Fig.~\ref{fig:eoverp}. 
The MC obtains the overall magnitude of signal and background well, 
although slight differences between MC and data based electron peaks are 
visible that are used to rescale the background fraction. The background 
level in the pion-enhanced sample ranges from 1 to 8\% for the different 
transverse-momentum bins and charges.

Using the extracted electron background fractions in pion-enhanced 
($r_\pi$) and electron-enhanced ($r_e$) samples, as well as the sample's 
respective asymmetries ($A_N^{Sig}$, $A_N^{BG}$), the charged-pion 
asymmetries $A_N^\pi$ can be obtained:

\begin{equation}
    A_N^\pi = \frac{r_e A_N^{Sig} - r_\pi A_N^{BG} }{r_e - r_\pi} \quad.
\end{equation}

The background asymmetries are consistent with zero, but their measured 
values were taken and the statistical errors propagated. The 
uncertainties on the signal to background fractions are assigned as 
systematic uncertainties as described below.

%-------------------------------------------- Fig_2

\section{Systematic studies and cross checks}

For the asymmetries, several methods were applied to ensure the validity 
of the results. Apart from the previously discussed square root formula 
(Eq.~\ref{eq:squareroot}), the single-spin asymmetries were also 
extracted by the so-called relative luminosity formula, where the 
relative luminosity $\mathcal{R} = \mathcal{L}^\uparrow / 
\mathcal{L}^\downarrow$ between the luminosities of the two spin 
orientations is applied to the count rate differences and the count rate sum:
\begin{equation}
    A_N = \frac{1}{P} \frac{ N_L^\uparrow - \mathcal{R} N_L^\downarrow}
    { N_L^\uparrow + \mathcal{R} N_L^\downarrow}\quad, 
\end{equation}
and normalized by the acceptance factor $\langle|\cos\phi|\rangle$. 
%The 
%relative luminosity accounts for the luminosity ratio between the two 
%spin states.

The asymmetries were additionally calculated as a function of the 
azimuthal angle and fitted with a sine modulation, although the 
azimuthal acceptance is limited at central rapidities in PHENIX. All 
three methods were compared using T-tests and were found to be 
consistent with each other such that no further systematic uncertainty 
was assigned.

The two polarized beams provide two independent measurements of the 
asymmetries at central rapidities and can therefore be compared with 
each other for consistency. Again using a T-test, the results were found 
to be consistent with each other, such that both results could be 
combined in the final result.

Another test is performed by randomizing the spin information of each of 
the beam crossings and fills to artificially remove any physical 
asymmetry. This so-called bunch-shuffling method is repeated many times 
and ideally should produce a Gaussian distribution centered at zero with 
a width as large as the statistical uncertainty. Any deviations would 
suggest additional systematic uncertainties that had not been included. 
While the bunch-shuffled asymmetries were on average consistent 
with zero, their widths were in some bins slightly larger than the 
statistical uncertainties. This variation was assigned as additional 
systematic uncertainty.

Lastly, the signal and background fractions that were described in the 
previous section were varied according to the uncertainties obtained 
from the fits to the $E/p$ distribution. These are the dominant 
systematic uncertainty, but the measurements are dominated by the 
statistical uncertainties for both charges and all transverse-momentum 
bins. Additionally, a global 3.4\% scale uncertainty exists due to the 
precision of the beam polarization evaluated by the RHIC polarimetry 
group~\cite{polarimetry}.

\section{Results}

\begin{figure}[htb]
\includegraphics[width=1.0\linewidth]{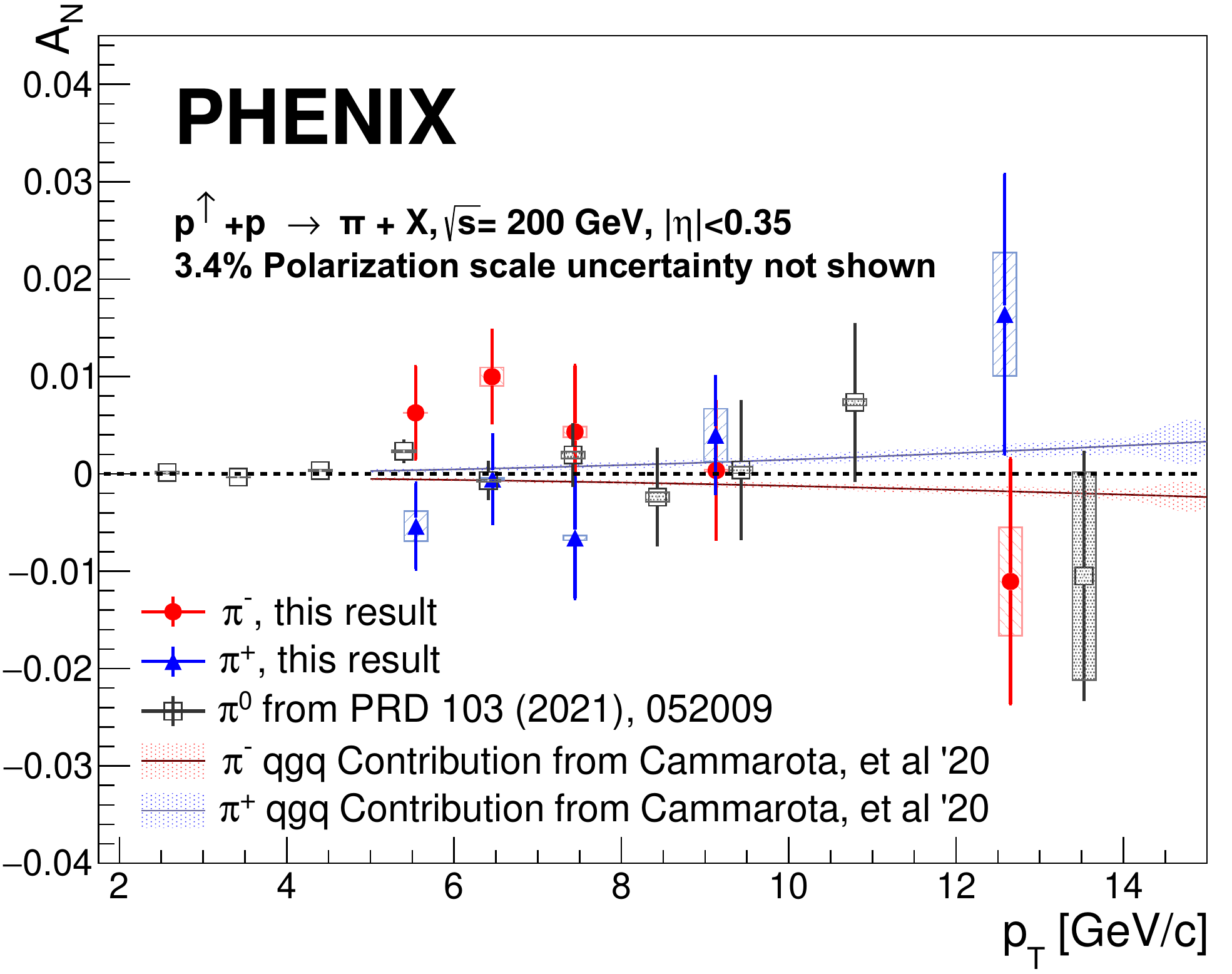}
\caption{Transverse-single-spin asymmetries as a function of transverse
momentum for positive (closed [blue] triangles), negative (closed [red]
circles), and the previously published neutral pions (open [black] boxes).
The statistical uncertainties are shown by bars and the systematic
uncertainties are shown by boxes. The lower [red] and upper [blue] curves
show the predicted $A_N$ and their uncertainties based
on Ref.~\protect\cite{Cammarota:2020qcw}.
}
\label{fig:AN}
\end{figure}

The asymmetries are displayed in Fig.~\ref{fig:AN} as a function of the 
transverse momentum and summarized in Table~\ref{tab:AN}. As can be 
seen, the results are statistically limited due to the fact that, in 
comparison to the neutral pions, only a fraction of charged pions shower 
in the EMCal and fire triggers.  Nevertheless, smaller than one percent level 
precision can be reached at lower transverse momenta, and the systematic 
uncertainties are generally much smaller.  While the asymmetries for 
each charge are consistent with zero, as well as consistent with the 
previously measured neutral-pion asymmetries, there are differences 
between positive and negative charges. The $\chi^2$ between $\pi^+$ and 
$\pi^-$ asymmetries is 9.04 for all five data points together. This 
might indicate a dependence of the asymmetries on the participating 
flavors of up and down quarks in particular, given that gluon-related 
initial-state effects would show the same behavior for either pion 
charge. Higher statistical precision, as is envisioned to be extracted 
with the sPHENIX experiment~\cite{sphenix}, will tell whether different 
flavors produce different asymmetries. It is worth noting that while the 
knowledge on the higher-twist correlations is still very limited, both 
the Sivers functions as well as the combinations of quark transversity 
and Collins fragmentation functions, show clear differences for up- and 
down-quark related effects.

When comparing the asymmetries to the expected asymmetries based on the 
global fit of data from SIDIS, other (predominantly forward single-spin 
asymmetries in $p$$+$$p$ collisions and $e^+e^-$ 
data~\cite{Cammarota:2020qcw}), the PHENIX results at higher transverse 
momenta are well described.  In the midrapidity range, these 
calculations are dominated by the higher twist quark-gluon-quark correlators 
in the nucleon that are related to the quark Sivers functions. However, at 
lower transverse momenta, the size and sign of the measured asymmetries, 
especially for negative pions, appear to be slightly different.

%-------------------------------------------- Fig_3

%=============================================== Table_I
\begin{table}[hbt]
\caption{Final single-spin asymmetries for charged pions as a function 
of transverse momentum with statistical errors ($\Delta A_N$) and 
systematic uncertainties ($\delta A_N$). An additional 3.4\% scaling 
uncertainty due to the beam polarization measurements is not shown.
}
\begin{ruledtabular} \begin{tabular}{ccccccc}
& $\pi^\pm$ & $p_T$ range & $A_N$ & $\Delta A_N$ & $\delta A_N$ & \\
&          & (GeV/$c$)  &     $(10^{-3})$ & $(10^{-3})$ & $(10^{-3})$ & \\ 
\hline
& $\pi^-$ 
   & 5--6    & 6.30  & 4.67  & 0.03  & \\
&   & 6--7    & 9.98  & 4.73  & 0.95 & \\
&   & 7--8    & 3.71  & 6.60  & 0.55 & \\
&   & 8--11   & 0.44  & 6.85  & 0.10 & \\
&   & 11--15  & -9.27 & 11.76 & 5.60 & \\
\\
&$\pi^+$ 
   & 5--6    & -5.60 &  4.33   & 1.55 &\\
&   & 6--7    & -0.44 &  4.49  & 0.10 & \\
&   & 7--8    & -6.81 &  6.12  & 0.23  &\\
&   & 8--11   &  3.99 &  5.91  & 2.70 & \\
&   & 11--15  & 15.43 & 11.83  & 6.33 & \\ 
\end{tabular} \end{ruledtabular}
\label{tab:AN}
   \end{table}

\section{Summary}

In summary, the PHENIX experiment has measured charged-pion 
transverse-single-spin asymmetries at central rapidities. The precision 
has been greatly improved from the previously published nonidentified 
charged-hadron asymmetries~\cite{PHENIX:2005jxc} to a one-percent-level 
precision of charged pions over a substantially larger range of 
transverse momenta. While the asymmetries overall are consistent with 
zero, as well as the previously published neutral-pion asymmetries in 
the same rapidity region, an indication for a charge-dependent 
separation of the asymmetries is visible. If confirmed with higher 
precision, it could indicate a flavor-dependent effect.

%%%%%%%%%%%%%%%%%%%%%%%%%  Acknowledgements 

%%%%%%%%%%%%%%%%%%%%%%  ACKNOWLEDGMENTS}  %%%%% MG21a version
%% 2018 change in Korea
%% 2021a dropped Brazil, Germany, and Pakistan -- no MGS qualifiers

\begin{acknowledgments}  % for PRC or PRD only 

We thank the staff of the Collider-Accelerator and Physics
Departments at Brookhaven National Laboratory and the staff of
the other PHENIX participating institutions for their vital
contributions.  
We acknowledge support from the Office of Nuclear Physics in the
Office of Science of the Department of Energy,
the National Science Foundation,
Abilene Christian University Research Council,
Research Foundation of SUNY, and
Dean of the College of Arts and Sciences, Vanderbilt University
(USA),
Ministry of Education, Culture, Sports, Science, and Technology
and the Japan Society for the Promotion of Science (Japan),
Natural Science Foundation of China (People's Republic of China),
Croatian Science Foundation and
Ministry of Science and Education (Croatia),
Ministry of Education, Youth and Sports (Czech Republic),
Centre National de la Recherche Scientifique, Commissariat
{\`a} l'{\'E}nergie Atomique, and Institut National de Physique
Nucl{\'e}aire et de Physique des Particules (France),
J. Bolyai Research Scholarship, EFOP, the New National Excellence
Program ({\'U}NKP), NKFIH, and OTKA (Hungary),
Department of Atomic Energy and Department of Science and Technology
(India),
Israel Science Foundation (Israel),
Basic Science Research and SRC(CENuM) Programs through NRF
funded by the Ministry of Education and the Ministry of
Science and ICT (Korea).
Ministry of Education and Science, Russian Academy of Sciences,
Federal Agency of Atomic Energy (Russia),
VR and Wallenberg Foundation (Sweden),
the U.S. Civilian Research and Development Foundation for the
Independent States of the Former Soviet Union,
the Hungarian American Enterprise Scholarship Fund,
the US-Hungarian Fulbright Foundation,
and the US-Israel Binational Science Foundation.

\end{acknowledgments}  % for PRC or PRD only

%%%%%%%%%%%%%%%%%%%%%%%%%%%  References 

%\bibliography{ppg247x1}   

 
\end{document}